\documentstyle[prl,floats,aps,twocolumn,multicol,epsf]{revtex}
\tighten
\begin{document}
\twocolumn[\hsize\textwidth\columnwidth\hsize\csname 
@twocolumnfalse\endcsname
\title{The singularity in supercritical collapse of a spherical scalar field}
\author{Lior M. Burko}
\address{
Department of Physics,
Technion---Israel Institute of Technology, 32000 Haifa, Israel}
\date{\today}

\maketitle

\begin{abstract}

We study the singularity created in the supercritical collapse 
of a spherical massless scalar field. We first model the geometry and the 
scalar field to be homogeneous, and find a generic solution 
(in terms of a formal series expansion) describing a 
spacelike singularity which is monotonic, scalar polynomial and strong. 
We confront the predictions of this analytical model with the 
pointwise behavior of fully-nonlinear and inhomogeneous numerical 
simulations, and find full compliance. We also study the phenomenology 
of the spatial structure of the singularity numerically. At 
asymptotically late advanced time the singularity approaches the 
Schwarzschild singularity, in addition to discrete points at finite 
advanced times, where the singularity is Schwarzschild-like. At other points 
the singularity is different from Schwarzschild due to the 
nonlinear scalar field. 
\newline
\newline
PACS number(s): 04.70.Bw, 04.20.Dw
\end{abstract}

\vspace{3ex}
]

\section{Introduction}

The spherically-symmetric collapse of a 
scalar field has been studied extensively in the last 
few years. It was shown that when one changes the value of a parameter which 
characterizes regular initial data (while freezing the other parameters),  
there exists a critical value of the parameter  
beyond which an apparent horizon 
(and consequently a black hole) forms. For parameters smaller than the 
critical ones, no stable configurations form, and the field 
disperses to infinity \cite{choptuik93,gundlach98}. 
Despite the significant advance gained recently in the 
understanding of dynamical gravitational collapse, the spacetime interior 
to the apparent horizon has received little attention. In particular, the 
nature of the spacetime singularity predicted to form inside the black 
hole in supercritical collapse, remained unresolved.
 
The linearized scalar-field perturbations of the interior of 
the Schwarzschild 
spacetime have been known for two decades now\cite{novikov78}. 
This leads naturally to the 
following questions: To what extent is the spacetime singularity in 
nonlinear collapse similar to the linearly perturbed Schwarzschild 
singularity? As the backreaction of the scalar field is 
expected to change the Schwarzschild singularity, what are the features 
and properties of this singularity? The goal of this paper is to study 
these questions within the model of the fully nonlinear collapse of a 
minimally-coupled, self-gravitating, spherically-symmetric, massless,   
real scalar field. 
We shall construct a simple analytical model to describe the 
singularity which evolves in this model, and confront it with 
fully-nonlinear numerical simulations. 

The organization of this paper is as follows. In 
Section \ref{secii} we present the analytical model and its main 
conclusions. We 
find a generic solution for the pointwise behavior at the singularity. 
This singularity is spacelike, scalar polynomial, and strong. In Section 
\ref{seciii} we describe the 
numerical simulations and compare these simulations with the predictions 
of the analytical model. We find full compliance between the analytic 
results  and the numerical simulations. We also study numerically the 
spatial structure of the singularity. We 
make some concluding remarks in Section \ref{seciv}.

\section{Analytical model}\label{secii}

The linear perturbation analysis of the interior of the Schwarzschild 
spacetime with a massless scalar field predicts the scalar field to diverge 
like $\ln r$ along radial $t={\rm const}$ lines \cite{novikov78}. 
Here, $r$ is the radial Schwarzschild coordinate  defined such that 
circles of radius $r$ have circumference $2\pi r$ ($r$ is timelike inside 
the black hole) and $t$ is normal to $r$ ($t$ is spacelike inside the black 
hole). (There is a gauge 
freedom in $t$---see below.) Therefore, near the singularity at $r=0$ one 
would no longer expect the linear solution to be valid. Rather, one would 
expect the backreaction on the metric to be significant. Consequently, in 
order to describe the singularity in this model one would need to solve the 
fully nonlinear Einstein-Klein-Gordon equations. Moreover, in the case of 
dynamical collapse the formation of the singularity is a nonlinear 
effect, and no linear perturbation analysis can be made in the first 
place. 

The analytical solution of the full Einstein-Klein-Gordon equations turns 
out to be very difficult even in spherical symmetry. However, the 
analysis of linearized perturbations of the Schwarzschild singularity 
\cite{novikov78} finds that at large values of $t$ the interior regions 
tend to a state which depends only on $r$. For any finite $r$ near the 
singularity the linear spherical scalar field vanishes like $t^{-3}$ at 
large $t$. 
One would expect, therefore, the deviations from Schwarzschild to be 
small at $t \to \infty$ (at finite $r$). It looks reasonable, therefore, 
to look for a simplified analytical model where dynamical variables 
depend only on $r$ also in the nonlinear case, at least for large $t$. 

Let us consider the general {\it homogeneous} spherically-symmetric 
line-element
\begin{equation}
\,ds^2=h(r)\,dt^2+f(r)\,dr^2+r^2\,d\Omega ^2 ,
\end{equation} 
where $\,d\Omega ^2=\,d\theta ^2+\sin ^2\theta\,d\phi ^2$ is the metric 
on the unit 2-sphere. The field equations are given by 
\begin{equation}
\left(f'r+f^2-f\right)/\left(fr^2\right)=\Phi '^2
\label{ei1}
\end{equation}
\begin{equation}
\left(h'r-hf+h\right)/\left(hr^2\right)=\Phi '^2
\label{ei2}
\end{equation}
\begin{eqnarray}
\left[\left(h^2\right)'\,\left(f-\frac{1}{2}rf'\right)+2rf\,\left(hh''
-\frac{1}{2}h'^2\right)-2f'h^2\right]\nonumber \\
/\left(4rfh^2\right)=-\Phi '^2
\label{ei3}
\end{eqnarray}
and by the Klein-Gordon equation $\nabla_{\mu}\nabla^{\mu}\Phi=0$ 
for the scalar field $\Phi$. Here, a prime denotes    
differentiation with respect to $r$. Because 
of the spherical symmetry and the homogeneity, the Klein-Gordon equation 
is readily integrated to 
\begin{equation}
\Phi '(r)=d\;\sqrt{\left|f/\left(hr^4\right)\right|} ,
\label{kg}
\end{equation}
where $d\ne 0$ is an integration constant, which will be shown below to be 
arbitrary, and to describe a pure gauge mode. We next use Eq.~(\ref{kg}) to  
eliminate the scalar field from Eqs.~(\ref{ei1})--(\ref{ei3}).  We obtain 
\begin{equation}
f'r+f^2-f+d^2\frac{f^2}{hr^2}=0
\label{e1} 
\end{equation}
\begin{equation}
h'r-hf+h+d^2\frac{f}{r^2}=0
\label{e2}
\end{equation}
\begin{equation}
h''-\frac{h'^2}{2h}+\frac{1}{r}\left(h'-h\frac{f'}{f}\right)-
\frac{h'f'}{2f}-2d^2\frac{f}{r^4}=0.
\label{e3}
\end{equation}
We assume a formal series expansion for $f,h$, and seek a generic 
solution for a spacelike singularity. We find that
\begin{eqnarray}
f(r)=&-&\left(\beta+1\right)Cr^{\beta +2}-
\frac{\left(\beta+1\right)^2\left(3\beta+4\right)}
{\left(\beta+2\right)^2}C^2r^{2\beta +4}\nonumber \\
&+&O\left(r^{3\beta+6}\right)
\label{f}
\end{eqnarray}
\begin{eqnarray}
h(r)=d^2Cr^{\beta}+d^2\frac{\beta\left(\beta+1\right)}
{\left(\beta+2\right)^2}C^2r^{2\beta+2}+O\left(r^{3\beta+4}\right) 
\label{h}
\end{eqnarray}
\begin{eqnarray}
\Phi (r)&=&\sqrt{\beta +1}\,\ln r+\frac{\sqrt{\beta +1}\,\left(
5\beta ^2+12\beta +6\right)}{\left(\beta+2\right)^3}Cr^{\beta+2}
\nonumber \\
&+&O\left(r^{2\beta+4}\right).
\label{p}
\end{eqnarray}
Here, $\beta \geq -1$ and $C>0$ are free parameters. This solution suggests 
the the complete formal series expansion is of the form
\begin{equation}
\left[\begin{array}{c}  f(r) \\ h(r)\end{array} \right] =
\left(\begin{array}{c}   1 \\ d^2/r^2\end{array} 
\right)\sum_{n=1}^{\infty}\left[\begin{array}{c} 
a_{n}^{f}\left(\beta ,C\right) \\ a_{n}^{h}\left(\beta ,C\right)
\end{array}\right]\,
r^{(\beta +2)n} . 
\label{series}
\end{equation}
It can be shown that the expansion coefficients $a_{n}^{f,h}$ can be 
found uniquely for any $n$ \cite{burko-97}. For the scalar field we find 
\begin{equation}
\Phi (r)=\sqrt{\beta +1}\,\ln r+\sum_{n=1}^{\infty}a_{n}^{\Phi}\left(
\beta, C\right)\,r^{(\beta +2) n}.
\label{p-series}
\end{equation}
We note that because the Klein-Gordon equation is linear in the scalar field 
$\Phi$, whereas the Einstein equations (\ref{ei1})--(\ref{ei3}) are 
quadratic in $\Phi$, the overall sign in Eq.~(\ref{p}) is arbitrary. 

We next discuss the genericity of the solution. The notion of a 
general solution for a system of nonlinear differential equations is not 
unambiguous. However, one can count the number of free parameters in the 
solution, and compare it with the expected number. If these two numbers 
are equal, than the solution is generic, in the sense that it captures a 
volume of non-zero measure in solution space. From the physical point of 
view one 
would expect two free parameters (one because of Birkhoff's theorem and 
one due to the scalar field). Apparently, our solution above has three 
free parameters, namely, $d^2, C, \beta$. However, the parameter $d^2$  
reflects a pure gauge mode. That is, one has the freedom to re-scale the 
coordinate $t$ by $t\to t'=T(t)$. In such a gauge transformation 
$d^2$ is changed (and in general the parameter $d^2$ can be 
replaced by any smooth function of $t$) without changing the physical 
content of the solution. (The parameter $d^2$ 
remains a constant parameter which does not depend on $t$ for the class 
of transformations $t\to t'=pt+q$ for constant $p$ and $q$.)  
Thus, the scalar field (\ref{p-series}) and curvature scalars 
below are independent of $d^2$. Consequently, our solution depends on two 
{\it physical} degrees of freedom ($\beta$ and $C$), and is therefore 
generic. 

The curvature can be described, say, by curvature scalars. We find that
to the leading order in $r$
\begin{equation}
R=-\frac{2}{C}\,r^{-(\beta +4)}
\label{ricci}
\end{equation}
\begin{equation}
R_{\mu\nu}R^{\mu\nu}=\frac{4}{C^2}\,r^{-(2\beta +8)}
\end{equation}
\begin{equation}
R_{\mu\nu\rho\sigma}R^{\mu\nu\rho\sigma}=
4\frac{2\beta^2+2\beta+3}{\left(\beta+1\right)^2C^2}\,r^{-(2\beta +8)}.
\end{equation}
Here, $R$ is the Ricci scalar, $R_{\mu\nu}$ is the Ricci tensor, and 
$R_{\mu\nu\rho\sigma}$ is the Riemann-Christoffel tensor. 
From these expressions the singularity is clearly scalar polynomial.  
We next show that it is strong in the Tipler sense 
\cite{tipler77}, namely, any extended physical object will unavoidably by 
crushed to zero volume upon hitting the singularity. Let us denote by 
$x^0$ the coordinate tangent to the worldline of an object infalling 
along a radial $t={\rm const}$ worldline, and by $\tau$ its proper time, 
set such that $\tau =0$ at $r=0$. Then, the $(00)$ tetrad component of 
the Ricci tensor, to the leading order in $\tau$, is given by 
\begin{equation}
R_{(0)(0)}(\tau )= 8\,\frac{\beta+1}{(\beta+4)^2}\,\tau^{-2}.
\label{ricci-tetrad}
\end{equation}
By Ref. \cite{clarke85}, Eq.~(\ref{ricci-tetrad}) implies that the 
singularity is strong in the Tipler sense, as 
\begin{equation}
\int_{\tau_{0}}^{\tau}\,d\tau '\int_{\tau_{0}}^{\tau '}\,d\tau ''
R_{(0)(0)}(\tau '')
\end{equation} 
diverges logarithmically at $\tau =0$. 

\begin{figure}
\input{epsf}
\centerline{\epsfysize 7.0cm
\epsfbox{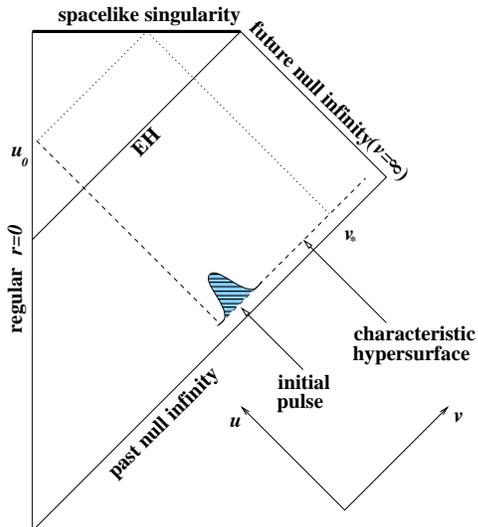}}
\caption{The Penrose diagram of the simulated spacetime. Prior to the 
characteristic hypersurface the geometry is Minkowskian. Due to the 
supercritical initial pulse a black hole is formed. For an outgoing null 
ray at $u_0$ the spacelike singularity is hit at $v=v_*$. Future null 
infinity is at $v=\infty$.} 
\label{fig0}
\end{figure}

Finally, let us define a `tortoise'
coordinate $r_*$ by $g_{r_*r_*}=-g_{tt}$. Then, null coordinates are
defined by $t=(v-u)/2$ and $r_*=(v+u)/2$, with ingoing $v$ and outgoing
$u$. For a certain outgoing ray $u=u_0\equiv \rm{const}$ which runs into the
spacelike singularity, the latter is hit at some finite value of
advanced time $v=v_*$ (see Fig.~\ref{fig0}). (Future null
infinity is located at $v=\infty$.)
We then find, to the leading order in $v_*-v$, that
\begin{equation}
r(u_0,v)=\left[d^2/(\beta +1)\right]^{1/4}\;(v_*-v)^{1/2}.
\label{r(v)}
\end{equation}

We note that as this model is homogeneous (the metric functions and the 
scalar field depend only on $r$), it actually describes just the pointwise 
behavior of the singularity. In general, the singularity is expected to 
depend not only on $r$ but also on $t$. This dependence on $t$ cannot be 
captured by our homogeneous model. The inhomogeneities will be studied 
therefore numerically. 

\section{Numerical simulations}\label{seciii}

\subsection{Initial data}

We constructed a numerical code based on double-null coordinates and on 
free evolution of the dynamical variables \cite{burko-ori97}. The code is 
stable and converges with second order (see below). We specify regular 
initial data on the characteristic hypersurface, which is located at 
$u=u_i$ and $v=v_i$ (see Fig.~\ref{fig0}). 
Prior to the initial pulse the geometry is Minkowskian. Then, due to the 
nonlinear scalar field, the geometry is changed, and an apparent horizon 
forms. At late advanced time ($v\to\infty$) the apparent horizon 
coincides with the event horizon (EH), and the black hole approaches 
asymptotically the (external) Schwarzschild black hole 
\cite{burko-ori97}. 

\begin{figure}
\input{epsf}
\centerline{\epsfysize 7.0cm
\epsfbox{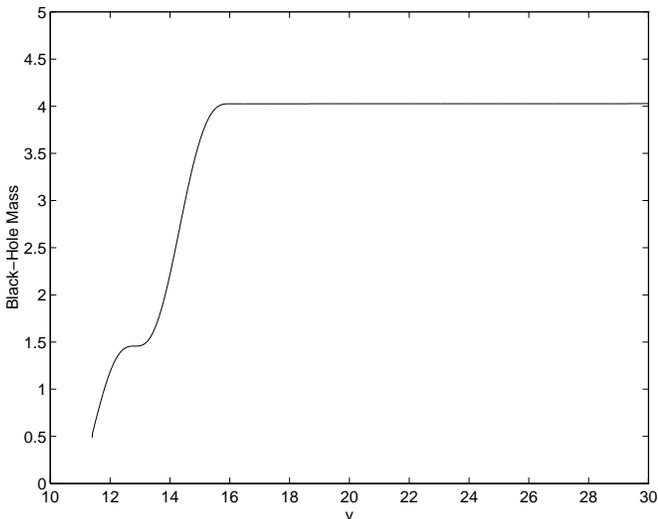}}
\caption{Black hole mass as function of advanced time $v$. The mass is 
evaluated by means of the value of $r$ at the apparent horizon.} 
\label{fig2}
\end{figure}

The details of the collapse model, the formulation of the 
characteristic problem, and the description of the code are given in 
Ref. \cite{burko-ori97}. Here, we shall describe them very briefly.  The 
null coordinates $u,v$ are defined such that $v=r$ and $u=(r-r_0)/r_{u0}$ 
on the characteristic hypersurface, where $r_0=r(u_i,v_i)$.  
(Note that this coordinate $v$ is not identical to the 
coordinate $v$ in Eq.~(\ref{r(v)}).)  
The initial data we specify are the following. The scalar 
field vanishes everywhere on the characteristic hypersurface, except for 
the compact interval $v_1<v<v_2$, where the scalar field has the form 
$\Phi (u_i,v)= A\sin ^2[\pi(v-v_1)/(v_2-v_1)]$. The rest of the initial 
data are chosen such that the constraint equations are satisfied on the 
characteristic hypersurface. The parameter which we vary is the amplitude 
$A$ in $\Phi (u_i,v)$. For small values of $A$ the initial data are 
subcritical, and no black hole forms. For large values of $A$, however, 
the initial data are supercritical, and a black hole forms. We shall 
discuss below only the latter case, namely, the internal structure of the 
black hole which forms in supercritical collapse. In the simulations 
described below the free parameters in the initial data are chosen to be  
$r_0=10$, $r_{u0}=-1/4$, $v_1=10$, $v_2=16$, and $A=0.29$. 
(These initial data correspond to a Minkowskian geometry for $v<v_1$ or 
$u<u_i$.)  These initial data result with a black hole with final mass 
$M_f\approx 4.02$. The value of $M_f$ can be easily determined from the 
value of $r$ at the apparent horizon: $r$ at the apparent horizon is just 
twice the mass of the black hole. Figure \ref{fig2} displays the mass of 
the black hole as a function of advanced time $v$. For $v<11.4$ the 
domain of integration does not intersect the apparent horizon, and 
therefore the mass of the black hole is not shown. (We note that because 
our code does not include the origin, we cannot find the apparent horizon 
of black holes with infinitesimal mass.) We find results similar to 
those reported below also 
for other values of supercritical initial-data parameters. (We have also 
studied the case of nonlinear perturbations of a pre-existing 
Schwarzschild black hole, by specifying the appropriate initial data. In 
this case we find that the properties of the singularity are similar to 
those found in supercritical collapse. In the latter case, however, the 
dynamical evolution of the spacetime geometry brings out the nonlinear 
aspects in a sharper way, and therefore we shall describe here only this 
latter case.) 

The only changes we introduced in the code which is described in Ref. 
\cite{burko-ori97} are related to
a careful approach to $r=0$: First, with uniform spacings of the grid in
advanced time one would hit the singularity within a finite integration
time and subsequently the code would crash. (However, one still needs to
preserve the dynamical refinement of the mesh in retarded time.) We
resolve this difficulty by checking the expected distance (in advanced time)
to the singularity on the last outgoing null ray of the grid for each
advanced-time step by means of the above analytical results (\ref{f}) 
and (\ref{h}). We then change the advanced-time step accordingly, such 
that crashing into the   
singularity is avoided, and in principle the code can run forever. (In
practice, because of the floating-point arithmetic limits of the machine
the approach to the singularity is limited. However, one can approach
$r=0$ sufficiently close to infer the asymptotic behavior from the
simulation.) Second, one would like to study the fields approaching 
different points along the singularity. Namely, for different values of 
$v_*$. This can be done in the following way. In Ref. \cite{burko-ori97} 
we introduce a procedure called ``chopping'', which is nothing but 
excision of the domain of integration slightly beyond the apparent 
horizon. Here, as we are interested in approaching the spacelike 
singularity at $r=0$, this procedure is performed only up to some finite 
value of advanced time $v_c$. The later $v_c$, the larger $v_*$, and the 
closer we arrive to the vertex point in the spacetime diagram \ref{fig0} 
which is at $v_*\to\infty$. This way we can arrive at different points 
along the spacelike singularity, and study the spatial dependence of the 
fields at the singularity. 

\subsection{Numerical simulations of the black hole interior}

We perform numerical simulations of the interior of the created black 
hole. Let us first  confront the predictions of the homogeneous model 
described in Sec. \ref{secii} with the pointwise behavior at the 
singularity as obtained from the fully-nonlinear and inhomogeneous 
numerical simulations. Then, we shall also study the phenomenology of the 
spatial structure of the singularity numerically. 

\subsubsection{The pointwise behavior}

\begin{figure}
\input{epsf}
\centerline{\epsfysize 7.0cm
\epsfbox{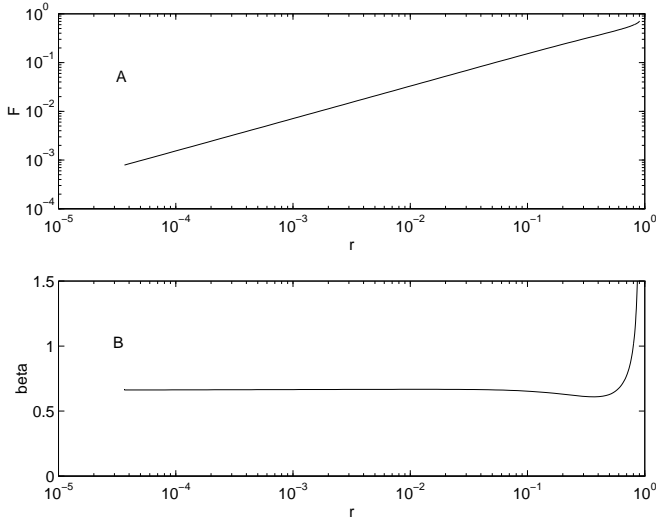}}
\caption{The behavior of the metric function $g_{uv}$ along an outgoing 
null ray: (A) $F=-2g_{uv}$ as a function of $r$. (B) the behavior of the 
local power index of $F$ as a function of $r$.} 
\label{fig3} 
\end{figure}

We first discuss the behavior of the metric functions near the 
singularity. In double null coordinates the metric function $g_{uv}$ is 
expected from Eq.~(\ref{h}) to behave near the singularity 
like $g_{uv}\propto r^{\beta}$ to leading order in $r$, because 
$g_{tt}=2(\,\partial u/\,\partial t)(\,\partial v/\,\partial t)g_{uv}$, 
and $\,\partial v/\,\partial t = 1= -\,\partial u/\,\partial t$. (The 
gauge freedom in $t$ is manifested just in a multiplicative factor, 
and not in the exponent $\beta$ which concerns us here.) 
Figure \ref{fig3} (A) displays $F=-2g_{uv}$ along an outgoing null ray at 
$u=u_0$, as a function of $r$. The 
straight line in Fig.~\ref{fig3} (A) indicates a power-law behavior. The 
asymptotic power-law behavior is checked more carefully in 
Fig.~\ref{fig3} (B), 
which shows the local power index of $F$ as a function of 
$r$. This local power index is nothing but the exponent 
$\beta$ in Eq.~(\ref{h}).  
As $\beta$ approaches asymptotically a constant value, the 
power-law behavior is verified. 

\begin{figure}
\input{epsf}
\centerline{\epsfysize 7.0cm
\epsfbox{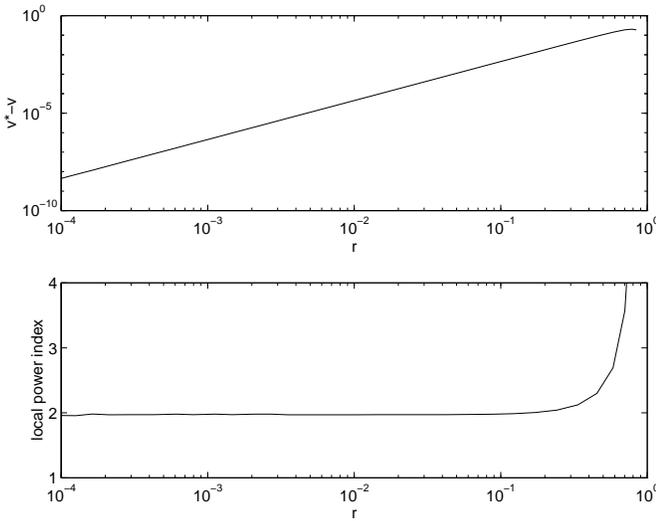}}
\caption{(A) $v_*-v$ as a function of $r$. (B) The local power index in 
(A) as a function of $r$.}
\label{fig4}
\end{figure}

The behavior of the metric function $r(u,v)$ is shown in Fig.~\ref{fig4} 
(A), which displays the distance from the singularity in 
terms of advanced time $v_*-v$, as a function of $r$ along the same 
outgoing null ray as in Fig.~\ref{fig3} (at $u=u_0$). Again, the straight 
line indicates a power-law behavior, whose local index is displayed in 
Fig.~\ref{fig4} (B). 
From Eq.~(\ref{r(v)}) this power-law index is expected 
to approach asymptotically a value of $2$. This is indeed found 
numerically, within a numerical error of $5\times 10^{-3}$. 

\begin{figure}
\input{epsf}  
\centerline{\epsfysize 7.0cm
\epsfbox{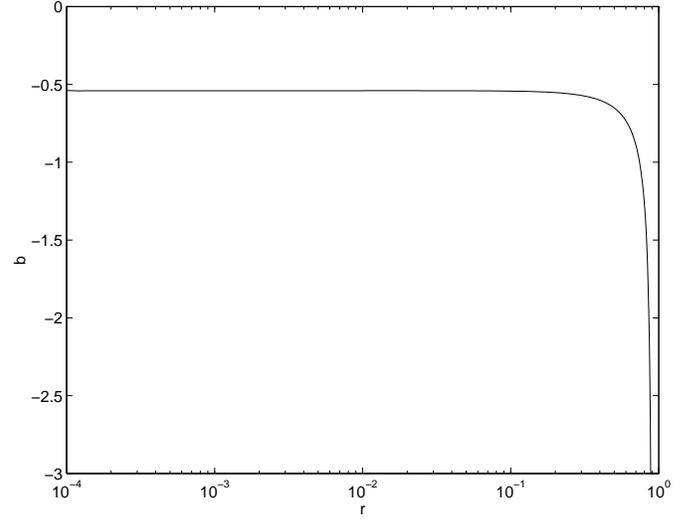}}
\caption{The value of $b=-1/(2r_{,u}r_{,v}r^2)$ as a function of $r$.}
\label{fig41}
\end{figure}

\begin{figure}
\input{epsf}
\centerline{\epsfysize 7.0cm
\epsfbox{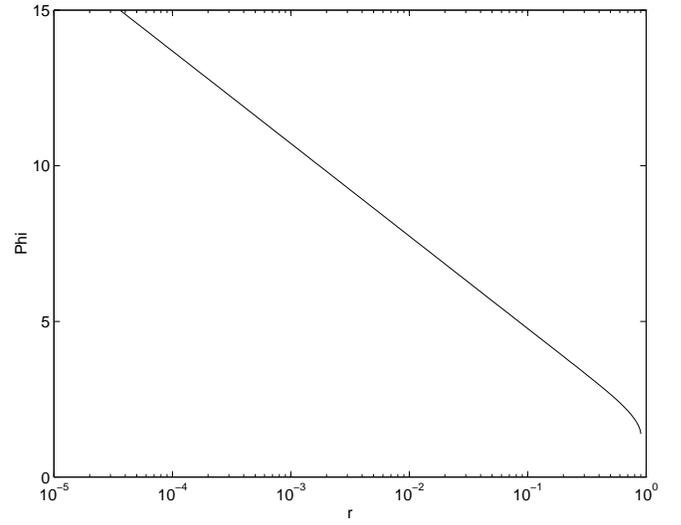}}
\caption{The scalar field $\Phi$ as a function of $r$.}
\label{fig5}
\end{figure}

Another prediction of the homogeneous model related to the metric
functions is that $g_{rr}/g_{uv}\propto r^2$, to the leading order in $r$.
Let us now find the analogue of this relation in the inhomogeneous case,
and compare the homogeneous and inhomogeneous cases numerically.
First, in Schwarzschild coordinates 
$g^{\mu\nu}\,\nabla_{\mu}r\,\nabla_{\nu}r$ is readily shown to equal 
$g^{rr}$. In double-null coordinates, 
$g^{\mu\nu}\,\nabla_{\mu}r\,\nabla_{\nu}r=
2g^{uv}r_{,u}r_{,v}=-(4/F)r_{,u}r_{,v}$, where $F(u,v)=-2g_{uv}(u,v)$. 
Combining these two expressions, one obtaines 
$g_{rr}(u,v)=-F/(4r_{,u}r_{,v})$. Consequently,
we find that $g_{rr}/g_{uv}=-1/(2r_{,u}r_{,v})$. Figure \ref{fig41} shows
$b=-1/(2r_{,u}r_{,v}r^2)$ along the same outgoing null ray at $u_0$, as a 
function of $r$. This ratio
approaches asymptotically a constant value, which confirms this
prediction of the homogeneous model. (The gauge freedom in $t$, which is
translated here into a gauge freedom in both $u$ and $v$, changes here
just the value of this constant.)

The behavior of the scalar field $\Phi$ along $u=u_0$ 
is shown in Fig.~\ref{fig5}. The 
straight line here indicates a logarithmic divergence of $\Phi$, as 
implied by Eq.~(\ref{p}). 

\begin{figure}
\input{epsf}
\centerline{\epsfysize 7.0cm
\epsfbox{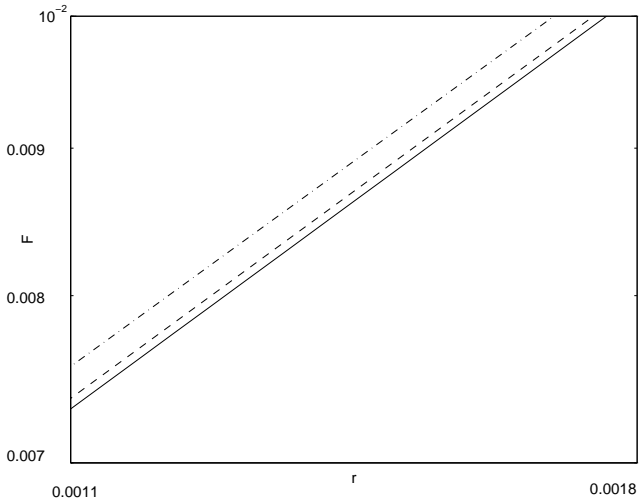}}
\caption{The metric function $F$ as a function of $r$ for three values of 
the grid parameter ($N=20,40,80$); Dashed-dotted: $N=20$, dashed: $N=40$, 
and solid line: $N=80$.} 
\label{fig6}
\end{figure}

\begin{figure}
\input{epsf}
\centerline{\epsfysize 7.0cm
\epsfbox{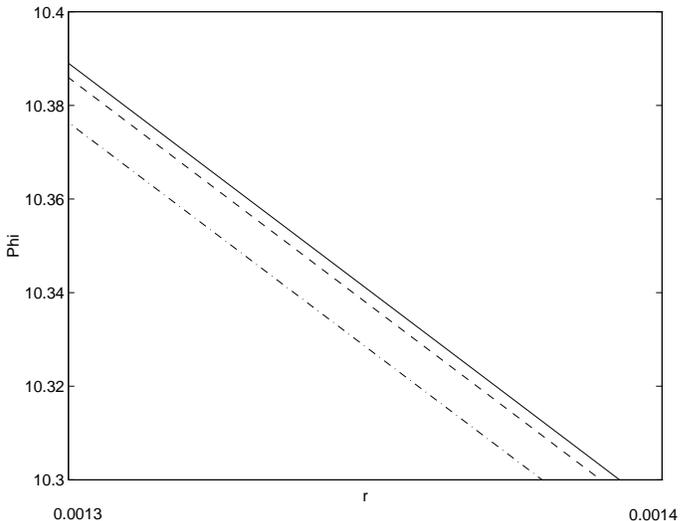}}
\caption{The scalar field $\Phi$ as a function of $r$ for three values of
the grid parameter ($N=20,40,80$); Dashed-dotted: $N=20$, dashed:
$N=40$, and solid line: $N=80$.}
\label{fig7}
\end{figure}

The stability and convergence of the code are demonstrated in Figs. 
\ref{fig6} and \ref{fig7}. Figure \ref{fig6} shows a magnified portion of 
Fig.~\ref{fig3} and Fig.~\ref{fig7} shows a magnified portion of Fig.~ 
\ref{fig5} for different values of the grid parameter $N$, namely, for 
$N=20,40$, and $80$, where $N$ is the (initial) number of grid points per 
a unit interval in both the ingoing and outgoing directions. The behavior in 
the two figures clearly indicates stability and convergence with second order. 

\begin{figure}
\input{epsf}
\centerline{\epsfysize 7.0cm
\epsfbox{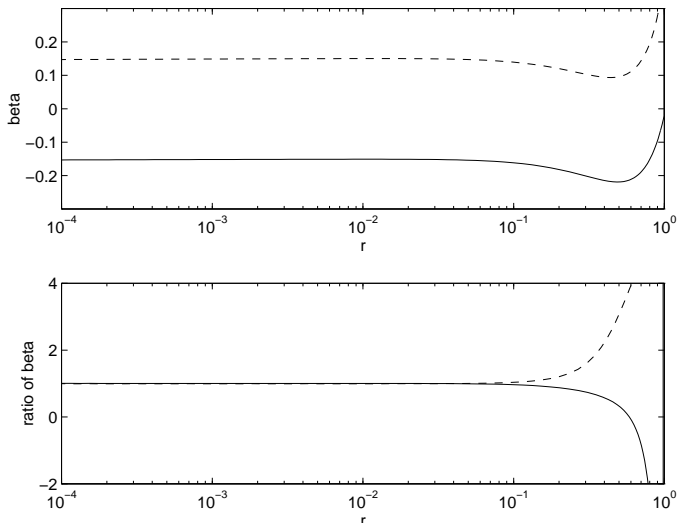}}
\caption{Top panel: The values of $\beta$ for two outgoing null rays as
functions of $r$. Bottom panel: The ratio of $\beta$ as determined from
the exponent in $g_{uv}$ and as determined from the amplitude of the
scalar field $\Phi$ for the same two outgoing null rays.}
\label{fig8}
\end{figure}

The homogeneous model also implies that $\beta$ appears in two places, 
which numerically are independent. Namely, in the exponent of $h(r)$ 
[given by Eq.~(\ref{h})] and in the amplitude of the scalar field 
[given by Eq.~(\ref{p})]. Figure \ref{fig8} shows the relation of these 
two values for $\beta$ for two differrent outgoing null rays: The top 
panel shows the values of $\beta$ for the two null rays as determined 
from Eq.~(\ref{h}) as functions of $r$, and the bottom panel the ratio of 
the two expressions for $\beta$ for the two outgoing rays, again as 
functions of $r$. We chose here the two outgoing null rays to be such 
that for one of them $\beta$ is positive and for the other $\beta$ is 
negative.  The lower panel in Fig.~\ref{fig8} indicates that the 
two expressions for $\beta$ are asymptotically 
equal to within one part in $10^{4}$. 

\begin{figure}
\input{epsf}
\centerline{\epsfysize 7.0cm
\epsfbox{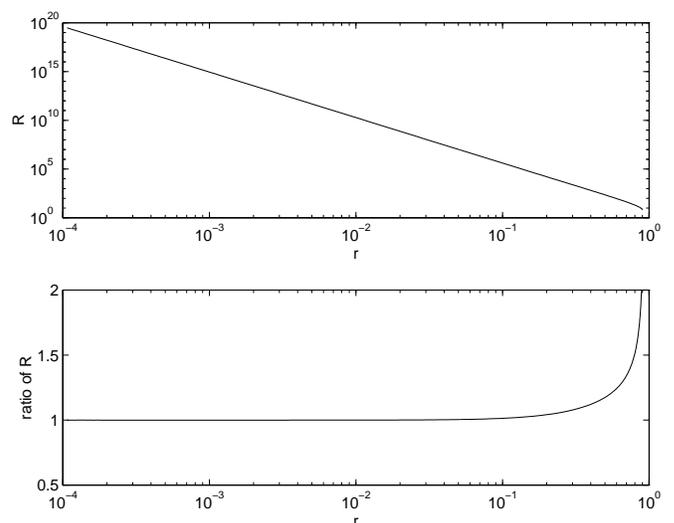}}
\caption{Top panel: The behavior of the Ricci scalar $R$ as
a function of $r$. Bottom panel: The ratio of $R$ as determined from
the fully nonlinear and inhomogeneous numerical simulations and the value
of $R$ as determined from the homogeneous model as a function of $r$.}
\label{fig9}
\end{figure}

The last prediction of the homogeneous model we test with the 
numerical simulations is the behavior of the Ricci curvature scalar given 
in Eq.~(\ref{ricci}). The top panel in Fig.~\ref{fig9} displays the 
behavior of the Ricci curvature scalar $R$ along the same outgoing null 
ray as the data in Figs. \ref{fig3}--\ref{fig7}, as a function of $r$. 
The value of $R$ can most easily be evaluated from the right hand side of 
the Einstein equations: The Einstein-Klein-Gordon equations are 
$R_{\mu\nu}=2\Phi_{,\mu}\Phi_{,\nu}$, and the Ricci scalar is simply 
given by $R=2g^{\mu\nu}\Phi_{,\mu}\Phi_{,\nu}$, which in double-null 
coordinates and spherical symmetry reduces to 
\begin{equation}
R=-\frac{8}{F}\Phi_{,u}\Phi_{,v}.
\label{r1}
\end{equation}
In order to evaluate the expression 
(\ref{ricci}) for $R$ we need first to express $C$. This can be done by 
comparing our result that $g_{rr}=-F/(4r_{,u}r_{,v})$ with  
the leading order in $r$ expression for $f$ from (\ref{f}). We 
find that to the leading order in $r$
\begin{equation}
C=\frac{1}{4}\,\frac{F}{\beta +1}\,\frac{1}{r_{,u}r_{,v}}
\,r^{-(\beta +2)}.
\label{c}
\end{equation}
The bottom panel of Fig.~\ref{fig9} shows the ratio of $R$ as calculated 
from Eq.~(\ref{r1}) and $R$ as calculated from Eqs.~(\ref{ricci}) and 
(\ref{c}), as a function of $r$. This ratio asymptotically deviates from 
unity by $3\times 10^{-4}$, despite the growth of $R$ by a factor of 
order $10^{20}$.

\subsubsection{The inhomogeneous behavior}

\begin{figure}
\input{epsf}
\centerline{\epsfysize 7.0cm
\epsfbox{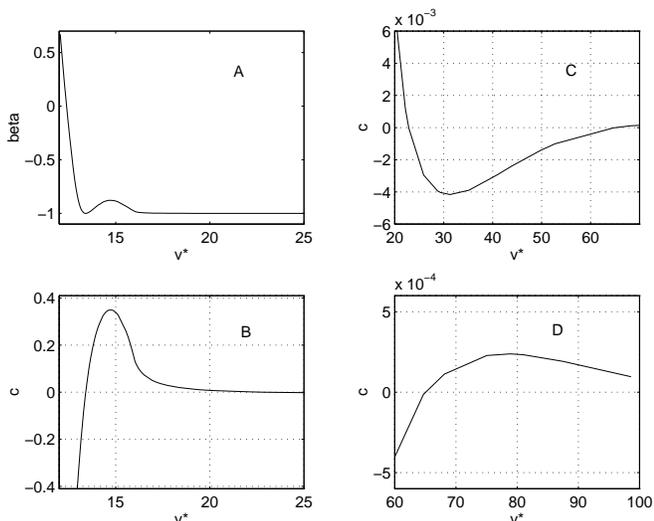}}
\caption{(A) $\beta$ as a function of $v_*$ along the spacelike
singularity. (B), (C), and (D): $c=r\Phi_{,r}$ as a function of $v_*$
along the spacelike singularity, for different sections of the latter.}
\label{fig10}   
\end{figure}

The inhomogeneous behavior near the singularity can be demonstrated by 
the variation of $\beta$ along the singularity. Namely, $\beta$ is no 
longer a parameter, but $\beta=\beta (t)$. [We would also expect 
$C=C(t)$.] Another issue which can be studied numerically is the 
overall 
sign issue of the scalar field. This sign is of importance, as its 
meaning is related to the divergence of $\Phi$ to $+\infty$ (for a minus 
sign) or to $-\infty$ (for a plus sign). We find that there are regions 
where $\Phi$ diverges to $+\infty$ and regions where $\Phi$ diverges to 
$-\infty$. Figure \ref{fig10} (A)  
shows $\beta$ as a function of $v_*$. For low values of $v_*$, $\beta$ is 
positive. Then, $\beta$ changes its sign, arrives at a minimum at 
which its value is $-1$, 
increases again, and finally, at late $v_*$, approaches asymptotically 
the value of $-1$. Although the numerical simulations indicate that 
$\beta$ has at least two more local minima, because of the numerical 
noise it is hard to locate them accurately. This can be done, however, in 
terms of the scalar field. 
The sign of the scalar field can be determined from 
$c=r\Phi_{,r}$, which, to the leading order in $r$ is $\mp\sqrt{\beta +1}$. 
Clearly, whenever $c(v_*)$ vanishes with a non-vanishing spatial gradient 
(namely, with $\,\partial c/\,\partial v_* \ne 0$), $\beta$ has a local 
minimum with a value of $-1$. From Figs. \ref{fig10} (B), (C), and (D) this 
happens at $v_*=13.4$, $22.5$, and $66.4$. We note that these values of 
$v_*$ are not the same as the values of $v$ along the EH where 
the quasinormal modes have their extrema or vanish. However, they may be 
related to the  quasinormal ringing, as their frequencies are similar. It 
is hard to quantify this and establish the hypothetical relation between 
the two oscillations, as we do not have enough cycles of the 
oscillations of $c$ on the singularity. Yet, Fig.~\ref{fig10}  
suggests that $c$ oscillates a number of times before it approaches a 
vanishing value for $v_*\to\infty$. In Fig.~\ref{fig10} (B), (C), and (D) 
the abscissae  
were chosen to overlap, such that one could infer the 
overall behavior. 

\begin{figure}
\input{epsf}
\centerline{\epsfysize 7.0cm
\epsfbox{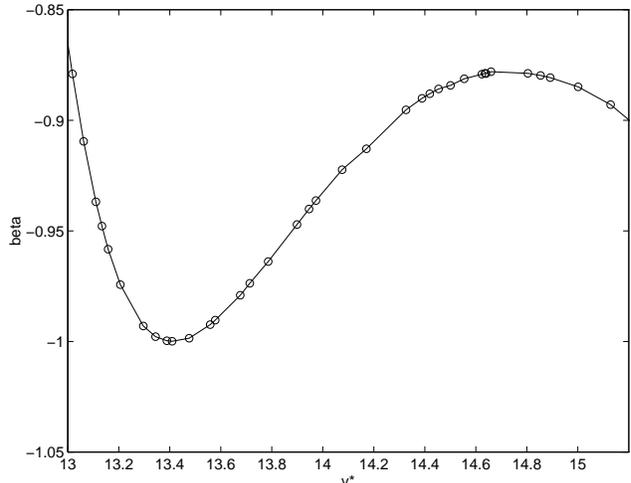}}
\caption{$\beta$ as a function of $v_*$. The circles are the data from
the numerical simulations, and the solid line is a linear interpolation
of these data.}
\label{fig11}   
\end{figure}

We find, that for a number of discrete points at finite values of $v_*$, 
$\beta =-1$. 
From Eq.~(\ref{p}) the scalar field $\Phi$ vanishes at these points. One 
could therefore ask, to what extent the singularity at these points 
is similar to the Schwarzschild singularity. Let us assume, that in the 
fully inhomogeneous case the leading order for the metric functions and 
the scalar field are given by the same form as the leading terms in the 
homogeneous case, except that the free parameters are replaced by 
functions of $t$. (Note that higher order terms will have a different 
form in the inhomogeneous case.) Then, in order that the singularity will 
be Schwarzschild-like at these points, both the scalar field and its 
gradient need to vanish there. The gradient of $\Phi$ will vanish only if 
$\,\partial\beta/\,\partial t$ vanishes. To see if indeed 
$\,\partial\beta/\,\partial t=0$ at the local minima of $\beta$ let us 
consider the neighborhood of the first minimum of $\beta (v_*)$. Figure 
\ref{fig11} shows $\beta$ as a function of $v_*$ near this first minimum. 
Figure \ref{fig11} suggests that indeed $\,\partial\beta/\,\partial t=0$ 
at the minimum, and consequently the singularity is 
expected to be Schwarzschild-like 
there, in the sense that at least the leading order for the scalar field 
and its gradient vanish there, and $\beta =-1$ as in Schwarzschild. Of 
course, one would need to consider also higher-order terms in order to 
study their contribution to the scalar field and to the geometry in an 
inhomogeneous model for a more detailed comparison of the singularity at 
these points and the Schwarzschild singularity.  

Another point to be made here is related to the limit $v_*\to\infty$. At 
this limit $\beta$ approaches a value of $-1$, and the spatial gradient 
of $\beta$ vanishes. This is evident from the behavior of $c$ for large 
$v_*$.  Consequently, the singularity asymptotically 
approaches the Schwarzschild singularity at late $v_*$. This conclusion 
agrees wih the conclusion drawn from the linear analysis in Ref. 
\cite{novikov78}. However, for any finite $v_*$ (except for the discrete 
points for which the scalar field and its gradient vanish on the 
singularity) $\beta\ne -1$. This means that 
the scalar field does not vanish on the singularity, and by its 
backreaction on the geometry the latter is not Schwarzschild. The 
deviation of the geometry from Schwarzschild is expected, however, to be 
small for small $\beta +1$, and decrease with growing $t$. In this sense 
the nonlinear numerical simulations confirm (at least in this model) 
the dictum of Doroshkevich 
and Novikov \cite{novikov78} that ``{\it a black hole has hair neither 
outside nor inside}.'' 

\section{Concluding remarks}\label{seciv}

We remarked above that in order to study the structure of the singularity 
in this model without the simplifying assumption of homogeneity, one 
could perhaps replace the free parameters in the solution with functions 
of $t$. If this were indeed the case, the leading terms are expected to 
retain their form. However, the higher-order terms in the solution will 
not. This is due to the appearance of non-vanishing spatial derivatives 
in the Einstein-Klein-Gordon equations. The solution of the full 
inhomogeneous equations looks like a formidable challenge even in the 
special case of spherical symmetry and a scalar field. 

Another point concerning the scalar field is related to its serving as a 
model for a dynamical physical field. The scalar field 
was introduced as a toy model, because it has a radiative mode in 
spherical symmetry. In the more realistic case of gravitational waves  the 
field does not have radiative modes in spherical symmetry, and therefore 
has no non-trivial 
dynamics. One could ask the following. To what extent do the spherical 
symmetry and the scalar field capture the essential dynamics of 
realistic black holes with vacuum perturbations, or the dynamical 
collapse of spinning fields? Realistic spinning black 
holes are known to have a Cauchy horizon, which is expected to transform 
into a null weak singularity \cite{ori92}. We have shown that in the 
model of the collapse of a spherical scalar field a spacelike singularity 
evolves. Will this be the case also in the collapse of gravitational 
waves? Although as yet there is no answer to this question, we could 
consider the model of the nonlinear perturbations of a spherical scalar 
field of a Reissner-Nordstr\"{o}m black hole. In this model one also has a 
Cauchy horizon in the unperturbed spacetime, and one can show that indeed 
a spacelike singularity evolves 
\cite{gnedin93,brady95,burko-97,burko97a,burko-ori98}. It turns out that 
there are 
some important differences between the spacelike singularity we find here 
(with vanishing electric charge) and the spacelike singularity in the 
model with charge, despite the similarity of the formal series 
expansions. First, the parameter $\beta$, which we find here to be greater 
than $-1$, is restricted to positive values in the presence of charge. 
Second, we find in the uncharged case that in the limit 
$v_*\to\infty$ the parameter $\beta\to -1$. In the charged case $\beta$ 
grows very rapidly for large $v_*$, and it is reasonable to expect it to 
diverge as $v_*\to\infty$. This difference in the asymptotic behavior of 
$\beta$ can perhaps be attributed to the following. In the uncharged case 
as $v_*\to\infty$ the outgoing Eddington-like coordinate 
$u_e\to -\infty$. However, because of the presence of the Cauchy horizon 
in the charged case, the limit $v_*\to\infty$ corresponds to a finite 
value of $u_e$. The infinite range of retarded time in the uncharged case 
acts to infinitely redshift the fields, an effect which does not have an 
analogue in the charged case.      

Finally, one might worry about the role that the scalar field plays in this 
model. Scalar fields are notorious for being related to peculiar 
phenomena. For example, scalar fields are known to destroy the 
oscillatory nature of the BKL singularity \cite{belinskii72}. Therefore, 
even though we find the singularity with a spherical 
scalar field (both in the uncharged and charged cases) to be monotonic, 
one perhaps could not infer 
from that on whether the (as yet hypothetical) spacelike singularity inside 
spinning black holes is monotonic or oscillatory.

\section*{Acknowledgements}

I thank Amos Ori for many helpful discussions and useful comments. 
This research was supported in part by the United States--Israel
Binational Science Foundation.

\end{document}